\documentclass[twocolumn,showpacs,preprintnumbers,amsmath,amssymb,superscriptaddress]{revtex4}
\usepackage{graphicx}
\usepackage{dcolumn}
\usepackage{color}
\usepackage{multirow}

\allowdisplaybreaks
\newcommand{\beq}{\begin{equation}}
\newcommand{\eeq}{\end{equation}}
\newcommand{\beqa}{\begin{eqnarray}}
\newcommand{\eeqa}{\end{eqnarray}}

\usepackage[colorlinks,plainpages=false,linkcolor=blue,urlcolor=blue,citecolor=blue,pdfpagemode=UseNone,pdfstartview=FitBH]{hyperref}

\begin{document}

\title{Chiral $d$-wave Superconductivity in a Triangular Surface Lattice Mediated by Long-range Interaction}

\author{Xiaodong~Cao}
\affiliation{Max-Planck-Institut f\"ur Festk\"orperforschung, Heisenbergstrasse 1, 70569 Stuttgart, Germany}

\author{Thomas~Ayral}
\affiliation{Physics and Astronomy Department, Rutgers University, Piscataway, NJ 08854, USA}
\affiliation{Institut de Physique Th\'{e}orique (IPhT), CEA, CNRS, UMR 3681, 91191 Gif-sur-Yvette, France }

\author{Zhicheng~Zhong}
\affiliation{Max-Planck-Institut f\"ur Festk\"orperforschung, Heisenbergstrasse 1, 70569 Stuttgart, Germany}
\affiliation{Ningbo Institute of Materials Technology and Engineering, Chinese Academy of Sciences, 315201 Ningbo, China}

\author{Olivier~Parcollet}
\affiliation{Institut de Physique Th\'{e}orique (IPhT), CEA, CNRS, UMR 3681, 91191 Gif-sur-Yvette, France }

\author{Dirk~Manske}
\affiliation{Max-Planck-Institut f\"ur Festk\"orperforschung, Heisenbergstrasse 1, 70569 Stuttgart, Germany}

\author{Philipp~Hansmann}
\affiliation{Max-Planck-Institut f\"ur Festk\"orperforschung, Heisenbergstrasse 1, 70569 Stuttgart, Germany}
\affiliation{Institut f\"ur Theoretische Physik, Eberhard Karls Universit\"at T\"ubingen, Auf der Morgenstelle 14, 72076 T\"ubingen}

\pacs{ }

\begin{abstract}
Correlated ad-atom systems on the Si(111) surface have recently attracted an
increased attention as strongly correlated systems with a rich phase
diagram. We study these materials by a single band model on the triangular
lattice including $1/r$ long-range interaction. Employing the recently
proposed TRILEX method we find an unconventional superconducting phase of
chiral $d$-wave symmetry in hole-doped systems. The superconductivity is
driven simultaneously by both charge \emph{and} spin fluctuations and is
strongly enhanced by the long-range tail of the interaction. We provide an
analysis of the relevant collective bosonic modes and explain how in
triangular symmetry both charge and spin channels contribute to the Cooper
pairing.
\end{abstract}
\date{\today}
\maketitle

The search for materials with unconventional high temperature
superconductivity (SC) has been one of the most active fields in correlated
solid state physics since the discovery of the cuprate high T$_{\rm{c}}$
compounds. Sophisticated synthesis technology nowadays allows for the
construction of new materials like heterostructures or surface systems on an
atomic length scale. Recently, many-body studies on experimentally well
controlled correlated ad-atom lattices X:Si(111) and X:Ge(111) with
(X=Pb,Sn,C) led to interesting results\cite{2010_PRB_Adatom_theory_Lechermann,
  2013_JPCM_Adatom_theory_Philipp, 2013_PRL_Adatom_theory_philipp,
  2013_NatComm_Adatom_theory_Gangli, hansmann2016uncertainty} and allowed to
unify the materials in a single phase
diagram~\cite{2013_PRL_Adatom_theory_philipp}. Due to sizable long-range
interaction in the triangular lattice geometry, some of the materials were
shown to be in close vicinity to a triple point between a Fermi liquid, a Mott
insulator, and a charge-ordered insulator. Sn:Si(111) and Pb:Si(111) in
particular turned out to be close to a charge-order Mott insulator phase
transition with sizable charge fluctuations visible in core level
spectroscopy~\cite{hansmann2016uncertainty} of Sn:Si(111). In complementary
studies \cite{2013_NatComm_Adatom_theory_Gangli} the importance of spin
fluctuations for Sn:Si(111) was emphasized. Such materials are, hence,
promising candidates to search for new physics like unconventional
superconductivity. 

For such systems theoretical methods are needed which are capable to capture
both local and non-local electronic correlations. Dynamical mean-field theory
(DMFT)~\cite{1996_Georges_rmp,KotliarRMP06} has been proven to be a powerful
approach to treat local correlations and Mott physics. If non-local
interactions have to be treated, extended DMFT
(EDMFT)~\cite{1996_PRL_edmft_Si} captures their effects on the local self
energy by a retarded onsite interaction. Local approximations like DMFT and
EDMFT are, however, not sufficient when non-local fluctuations start to play
an important role. To overcome these shortcomings of DMFT, several extensions
have been proposed~\cite{rohringer2017diagrammatic,Maier2005a}. Cluster
extensions of DMFT in real and reciprocal
space~\cite{Hettler1998,Lichtenstein2000,Kotliar2001,Maier2005a}, e.g., are
capable to treat non-local short range fluctuations. Long range fluctuations,
on the other hand, can be taken into account by
DMFT+GW~\cite{2002_PRB_edmft_Sun,2004_PRL_edmft_Sun,2013_PRB_edmft_thomas} or
dual boson
methods~\cite{2012_AOP_DualBoson_Lichtenstein,2014_PRL_DualBoson_Loon,2016_PRB_DualBoson_Loon,2014_PRB_DualBoson_Hafermann}. For
our study we employ the recently developed TRILEX
approximation~\cite{2015_PRB_trilex_thomas,
  2016_PRB_trilex_thomas,2017_thomas_sc_trilex, 2017_cluster_trilex_thomas}
which combines a balanced treatment of long range spin and charge fluctuations
with comparatively little computational effort.

In this letter we show that the triangular lattice model for the ad-atom
materials has a dome shaped superconducting phase of chiral $d$-wave symmetry
as a function of hole doping in realistic parameter regimes. The long-range
interaction is key for enhanced critical temperatures and distinguishes the
ad-atom Hamiltonian from triangular Hubbard
models~\cite{2008_zhou_prl,2008_su_prb,2010_Kuroki_prb,2012_nandkishore_chiral,2013_PRB_triangular_SC_Juana,
  2012_Platt_prb, 2013_platt_prl, 2014_Annica_prb}. By analyzing spin- and
charge response functions we further show that the pairing mechanism crosses
over from a cumulative spin/charge fluctuation character at small dopings to
a charge dominated one at large doping.\\

The low energy Hamiltonian on the triangular lattice with long-range interaction reads:
\begin{gather}\label{Hamiltonian}
 H = \sum_{i,j,\sigma}t_{ij}\hat{c}^{\dagger}_{i\sigma}\hat{c}_{j\sigma} + \frac{1}{2}\sum_{i,j}U_{ij} \hat{n}_{i} \hat{n}_{j} - \mu\sum_{i}\hat{n}_{i},
\end{gather}
where $\hat{c}^{\dagger}_{i\sigma}$ ($\hat{c}_{i\sigma}$) are electron
creation (annihilation) operators on site $i$ with spin
$\sigma=\uparrow,\downarrow$. $\hat{n}_{i}=\hat{n}_{i\uparrow}+\hat{n}_{i\downarrow}$
is the density operator on site $i$, and $\mu$ is the chemical
potential. $t_{ij}$ and $U_{ij}$ are the hopping integrals and long-range
Coulomb interaction strength between sites $i$ and $j$. For translational
invariant two-dimensional systems, the long-range Coulomb interaction, in
momentum space, reads $U_\mathbf{q} = U_{0}+V\sum_{i\neq 0} e^{i\mathbf{q}
  \cdot \mathbf{R}_{i}}/|\mathbf{R}_{i}|$ where $\mathbf{R}_{i}$ are real
space coordinates, $U_{0}$ is the on-site interaction, and $V$ is the strength of the long-range interaction respectively (Suppl. Mat. A). More specifically, we adopt hopping parameters up
to next-nearest-neighbors ($t=0.042$eV and $t'=-0.02$eV) from
~\cite{2013_JPCM_Adatom_theory_Philipp,2013_PRL_Adatom_theory_philipp} derived
from density functional theory (DFT) for the Pb:Si(111) ad-atom system
(closest to the triple point) and vary the interaction parameters in realistic
regimes for the ad-atom materials found by constrained random phase
approximation ~\cite{2013_PRL_Adatom_theory_philipp}. 

\begin{figure}[t]
  \includegraphics[width=\columnwidth]{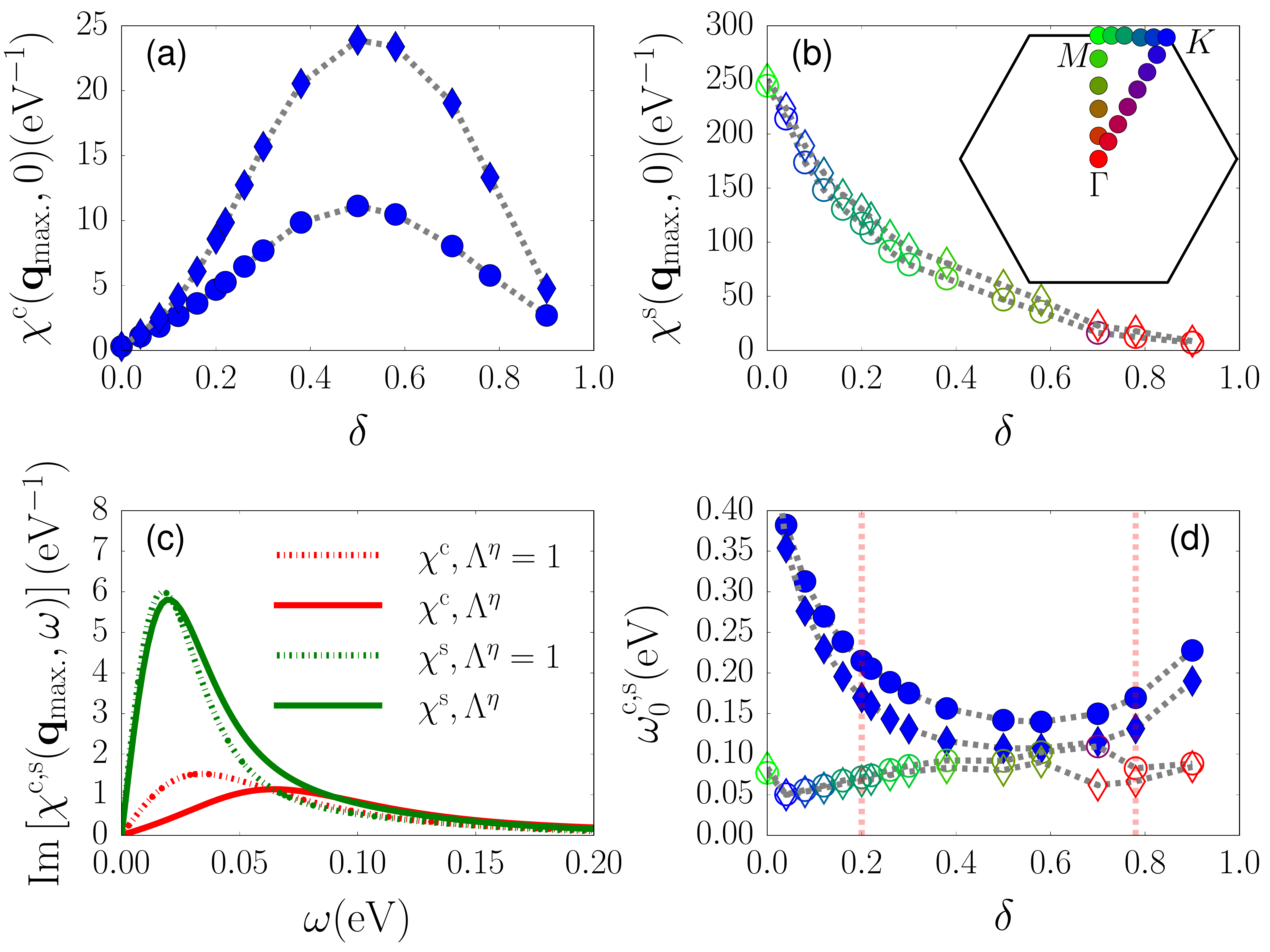}
  \caption{Maximum values of the static charge (a) and spin (b) response
    functions versus hole doping. Color coding indicates the position of the
    maximum in the first Brillouin zone as defined in the inset. Data is shown
    for fixed $U_{0}=0.7$eV and $T=40$K and non-local interaction strength $V=0.3$eV (diamonds) and $V=0.2$eV (circles); (c) Charge- and spin response functions on the real frequency
    axis (obtained by analytical continuation with the maximum entropy
    method\cite{JARRELL1996133}) at their maximum in momentum space
    ($\mathbf{q}_{\rm{max.}}$) with (dashed) and without (solid) vertex corrections
    for $T=116$K and $\delta=0.2$; (d) Characteristic frequency of charge-
    (filled symbols) and spin (open symbols) fluctuations with the same
    convention and parameters as (a) and (b).}
\label{Fluctuations}
\end{figure}

TRILEX approximates the three-legged fermion-boson interaction vertex using a
local self-consistent quantum impurity model. For systems retaining
$\rm{SU(2)}$ symmetry, the self-consistent TRILEX
equations~\cite{2015_PRB_trilex_thomas,
  2016_PRB_trilex_thomas,2017_thomas_sc_trilex, 2017_cluster_trilex_thomas}
for the fermionic single particle self-energy $\Sigma(\mathbf{k},i\omega_{n})$
and bosonic polarization in charge and spin channel
$P^{\rm{c,s}}(\mathbf{q},i\nu_{n})$ can be rewritten as:
\begin{align}\label{trilex_self_consistensy_equation}
\Sigma_{\mathbf{k}, i\omega_{n}} = \nonumber\Sigma^{\mathrm{imp}}_{i\omega_{n}} -\!\! \sum_{\eta,\mathbf{q},i\nu_{n}}m^{\eta}\widetilde{G}_{\mathbf{k}+\mathbf{q},i\omega_{n}+i\nu_{n}}\widetilde{W}^{\eta}_{\mathbf{q},i\nu_{n}}\Lambda^{\mathrm{imp},\eta}_{i\omega_{n},i\nu_{n}}\\
P^{\eta}_{\mathbf{q},i\nu_{n}} =P^{\mathrm{imp},\eta}_{i\nu_{n}} +2\sum_{\mathbf{k},i\omega_{n}}\widetilde{G}_{\mathbf{k}+\mathbf{q},i\omega_{n}+i\nu_{n}}\widetilde{G}_{\mathbf{k},i\omega_{n}}\Lambda^{\mathrm{imp}, \eta}_{i\omega_{n},i\nu_{n}}
\end{align}
where the index $\eta=\rm{\{c,s\}}$ corresponds to charge and spin channel
respectively, and $\omega_{n}$ and $\nu_{n}$ are fermionic and bosonic
Matsubara frequencies. $G_{\mathbf{k},i\omega_{n}}$ is the dressed Green's
function, and $W^{\rm{c,s}}_{\mathbf{q},i\nu_{n}}$ are the fully screened
interactions in the charge and spin channel respectively. The local part of
self-energy and polarization are replaced by their impurity counterparts
$\Sigma^{\rm{imp}}_{i\omega_{n}}$ and $P^{\rm{imp},\eta}_{i\nu_{n}}$
respectively, and for any quantity $X$,
$\widetilde{X}_{\mathbf{k},i\omega_{n}}=X_{\mathbf{k},i\omega_{n}}-X^{\mathrm{loc}}_{i\omega_{n}}$
with $X^{\mathrm{loc}}_{i\omega_{n}}=\frac{1}{N_{k}}\sum_{\mathbf{k}\in
  \mathbf{B.Z.}}X_{\mathbf{k},i\omega_{n}}$. We employ the Heisenberg
decomposition of the interaction \cite{2016_PRB_trilex_thomas}, for which we
have $m_{\rm{c}}=1$, $m_{\rm{s}}=3$ and $W^{\eta}_{\mathbf{q},i\nu_{n} } =
U^{\eta}_{\mathbf{q}}
\left[1-U^{\eta}_{\mathbf{q}}P^{\eta}_{\mathbf{q},i\nu_{n}}\right]^{-1}$. Bare
interactions in charge and spin channel are, hence, given by
$U^{\rm{c}}_{\mathbf{q}} = \frac{U_{0}}{2}+v_{\mathbf{q}}$ and $U^{\rm{s}} =
-\frac{U_{0}}{6}$. This spin/charge ratio is a choice (dubbed \lq\lq Fierz
ambiguity" \cite{2016_PRB_trilex_thomas,
  2017_cluster_trilex_thomas}). Moreover, in the parameter range explored in
this paper we have observed (Fig.~\ref{Fluctuations} and
Suppl. Mat. B) that using
$\Lambda^{\mathrm{imp},\eta}_{i\omega_{n},i\nu_{n}}\approx 1$ in
Eq.~\eqref{trilex_self_consistensy_equation} does not change our results
qualitatively as it was also found in~\cite{2017_thomas_sc_trilex}. This
simplified TRILEX version can be seen as a GW+EDMFT like scheme which,
however, can treat simultaneously both charge and spin fluctuations. The
impurity problem was solved using the segment picture in the
hybridization-expansion continuous time quantum Monte-Carlo
algorithm~\cite{ctqmc,ctqmc_seg_1,ctqmc_seg_2,ctqmc_seg_3,ctqmc_seg_4}
implemented with the TRIQS library~\cite{triqs}. 

In order to probe superconductivity instabilities, we solve the linearized gap
equation with converged simplified TRILEX results as an
input~\cite{2017_thomas_sc_trilex}. For singlet $d-$wave pairing the
corresponding eigenvalue equation for the gap reads
\begin{equation}\label{Gap}
  \lambda\Delta_{\mathbf{k},i\omega_{n}} = -\sum_{\mathbf{k}',i\omega_{n}'}|G_{\mathbf{k}',i\omega_{n}'}|^{2} \Delta_{\mathbf{k}',i\omega_{n}'} V^{\rm{eff}}_{\mathbf{k}-\mathbf{k}',i\omega_{n}-i\omega_{n}'},
\end{equation}
where the singlet pairing interaction is given by 
\begin{equation}
V^{\rm{eff}}_{\mathbf{q},i\nu_{n}} = m^{\rm{c}}W^{\rm{c}}_{\mathbf{q},i\nu_{n}} -m^{\rm{s}}W^{\rm{s}}_{\mathbf{q},i\nu_{n}}
\end{equation}
and is therefore a combination of effective interaction in \emph{charge and
  spin} channel. The SC instability occurs when the largest eigenvalue $\lambda
= 1$. The pairing symmetry is monitored by the $\mathbf{k}$ dependence of the
gap function $\Delta_{\mathbf{k},i\omega_{n}}$.\\

\begin{figure}[t]
  \includegraphics[width=\columnwidth]{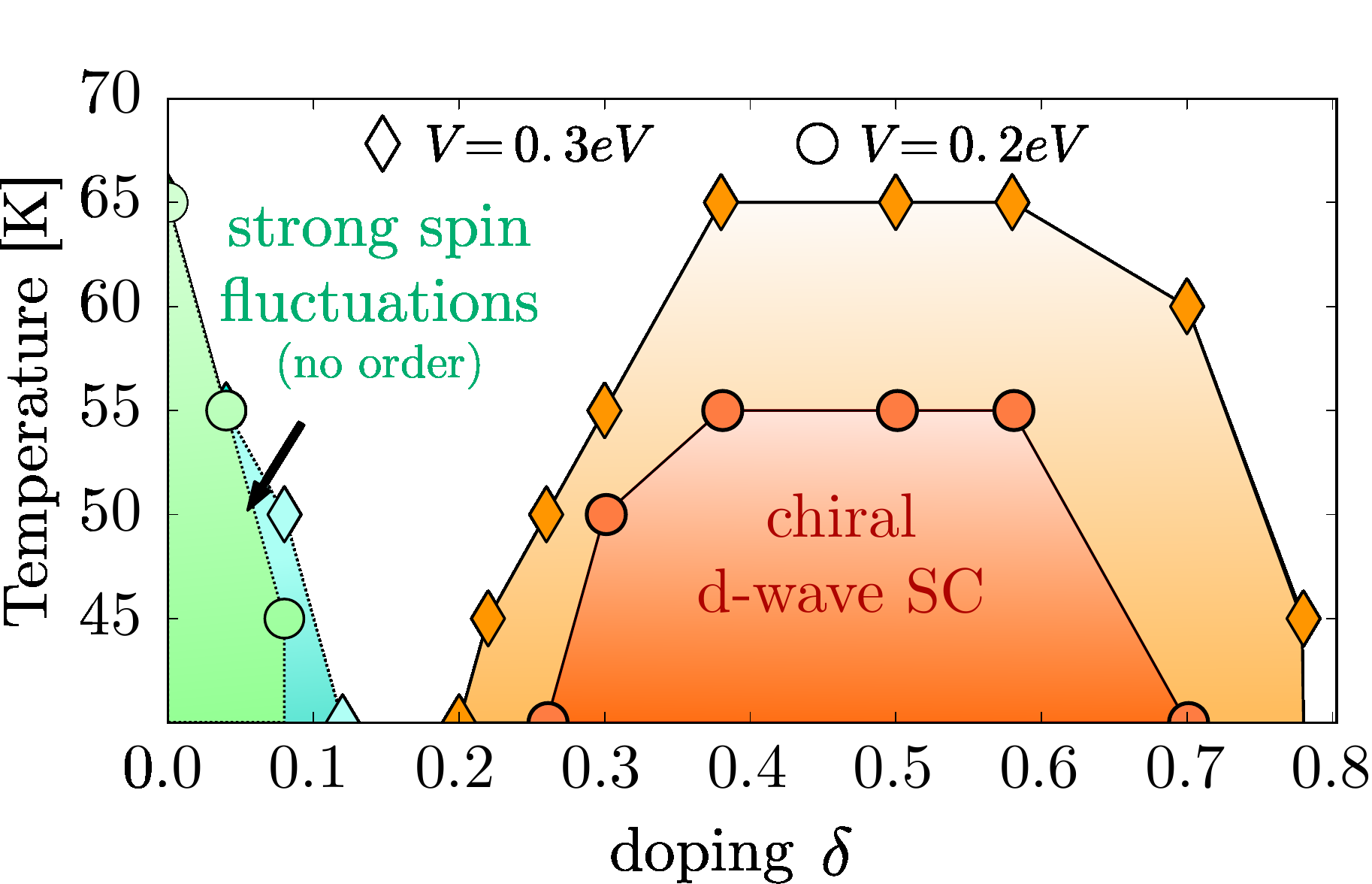}
  \caption{Phase diagram of the Hamiltonian ~\eqref{Hamiltonian} as function
    of temperature (for $T>40$K) and doping for $U_{0}=0.7$eV, $V=0.2$eV
    (circles) and $V=0.3$eV (diamonds). Green/blue regions correspond to
    $1\geqslant\rm{Max}{\left[-P^{\rm{s}}(\mathbf{q},i\nu_{n}=0)U^{\rm{s}}\right]}
    \geqslant 0.95$ for $\mathbf{q}\in\mathbf{B.Z.}$. Orange/red regions indicate chiral $d-$wave
    superconductivity.} 
\label{PhaseDiagram}
\end{figure}

\textit{Emergence of d-wave superconductivity --}
In Fig.~\ref{PhaseDiagram} we plot the temperature--doping ($T$--$\delta$)
phase diagram for $V=0.2$eV and $V=0.3$eV for a fixed value of $U_{0}=0.7$eV
in the simplified TRILEX approximation. At half-filling ($\delta=0$) we obtain
a correlated Fermi liquid (Suppl. Mat. C) with strong magnetic
fluctuations. The static spin-spin correlation function
$\chi^{\rm{s}}(\mathbf{q},i\nu_{n}=0)$ is very large at some $\mathbf{q}$ but
has not diverged yet, i.e. no phase transition has occurred. More precisely,
we use $\rm{Max}{\left[-P^{\rm{s}}(\mathbf{q},i\nu_{n}=0)U^{\rm{s}}\right]}$
with $\mathbf{q}\in\rm{B.Z.}$ which reaches $1$ at a second order spin ordering
phase transition to quantify the strength of the spin fluctuations and color
code regions in the phase diagram for which
$1>\rm{Max}{\left[-P^{\rm{s}}(\mathbf{q},i\nu_{n}=0)U^{\rm{s}}\right]}
\geqslant 0.95$ in green ($V=0.2$eV) and blue ($V=0.3$eV). From this
plot we see that spin fluctuations are slightly enhanced by increasing
$V$. For $\delta>0.2$ we observe the emergence of a dome-shaped
superconducting phase (a plot of the $\lambda$ parameter in Eq.~\eqref{Gap} as
a function of temperature is shown in the Suppl. Mat. D). The
pairing symmetry of the SC phase is of $d$-wave character and includes doubly
degenerate $d_{x^{2}-y^{2}}$- and $d_{xy}$-wave pairing channels (see Suppl. Mat. E for a plot of the gap
function). The degeneracy of these two pairing symmetries is protected by the
C$_{6v}$ point group of the triangular lattice, which then yields chiral
$d-$wave symmetry below T$_{\rm{c}}$ to maximize condensation energy. The predicted
chiral SC phase depends crucially on $V$: T$_{\rm{c}}$ increases from $V=0.2$eV (red circles) to $V=0.3$eV (orange diamonds) as shown in Fig.~\ref{PhaseDiagram}. Moreover,
for $V=0.0$eV and $V=0.1$eV (not shown here) we do not find a SC phase for
$T>40$K. 

\begin{figure}[t]
  \includegraphics[width=\columnwidth]{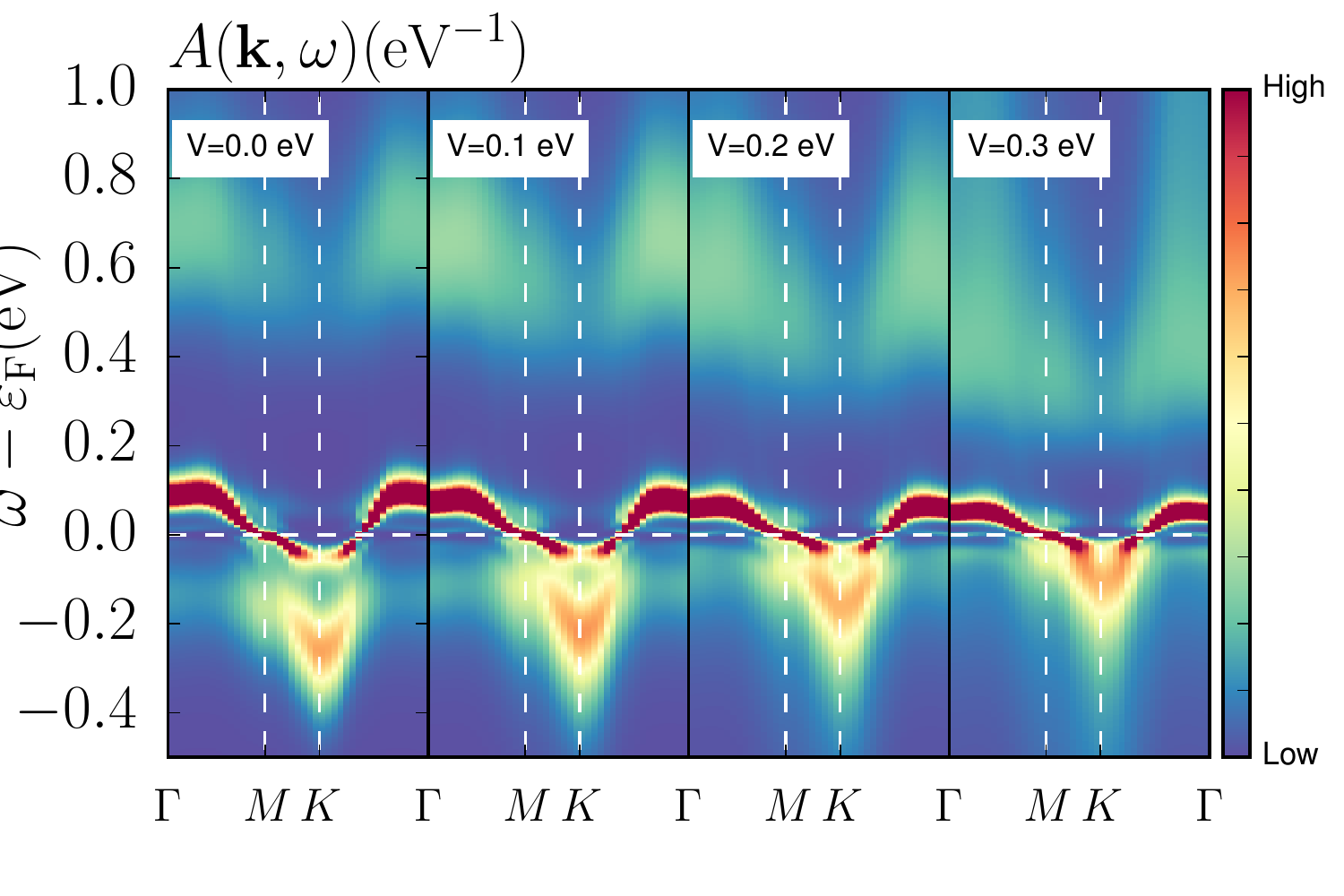}
  \caption{Single particle spectral function $A(\mathbf{k},\omega)$ along the
    path $\Gamma$-M-K-$\Gamma$ (see inset of Fig.~\ref{Fluctuations}) for
    fixed doping $\delta=0.2$, $T=40$K, $U_0=0.7$eV and four values of $V$.}
\label{OneparticleSpectal}
\end{figure}

\textit{Impact of long-range interaction on susceptibilities and single particle spectra --}
The impact of $V$ on the SC instability is reflected in the effective
singlet-pairing interaction $V^{\rm{eff}}_{\mathbf{q},i\nu_{n}}$ which depends
on fluctuations in \emph{both} charge and spin channels. We analyze the
respective susceptibilities $\chi^{\rm{c}/\rm{s}}(\mathbf{q},i\nu_{n})$ with
the data shown in Fig.~\ref{Fluctuations}: In the upper panels we show the
maximum values of the static ($i\nu_{n}=0$) charge (left hand side) and spin
(right hand side) susceptibilities as a function of hole doping. The
corresponding position of the maximum in the first Brillouin zone is color
coded (see inset).

The charge fluctuations increase with hole doping to a maximum value around
$\delta=0.5$ and, thereafter, decrease approaching the ``empty'' limit at
$\delta=1$. The spin fluctuations, instead, decrease monotonically as a
function of $\delta$. While $\chi^{\rm{c}}(\mathbf{q},i\nu_{n}=0)$ always
peaks at $K$, the maximum of $\chi^{\rm{s}}(\mathbf{q},i\nu_{n}=0)$ moves from
$M$ to $K$ when the system is slightly doped, and then follows $K\rightarrow
M\rightarrow \Gamma$ when the system is further hole-doped. The peak position
of the charge response function as a function of doping remains at the $K$ point since
its momentum dependence is mainly determined by the doping independent
$v(\mathbf{q})$ which energetically favors a $3\times 3$ charge configuration
in real space~\cite{2013_PRL_Adatom_theory_philipp}. The momentum dependence
of the spin response function, however, is mostly determined by the topology
of the Fermi surface. Indeed, the $V$ dependence is much stronger for the
charge response (compare diamond ($V=0.3$eV) and circle ($V=0.2$eV) symbols in
Fig.~\ref{Fluctuations}). There are, however, small effects of $V$ to the spin
response function which can be understood by the $V$-dependent renormalization
of the one-particle spectra as show in
Fig.~\ref{OneparticleSpectal}~\cite{2016_JPCM_Werner}. At fixed $T=40$K and
$\delta=0.2$, $V$ is increased from $0.0$eV to $0.3$eV (subplots from left to
right hand side). Upon increasing $V$, the bandwidth is effectively reduced
and the spectral weight near to the Fermi energy is increased. Consequently,
particle-hole excitations that contribute to the spin polarization
$P^{\rm{s}}(\mathbf{q},i\nu_{n})$ and the spin susceptibility are enhanced.

We now extend these considerations to the frequency dependence of the bosonic
fluctuations. In Fig.~\ref{Fluctuations}c we plot the dynamic response
functions at the $\mathbf{q}$-points where they are maximal
($\mathbf{q}_{\text{max.}}$) for doping $\delta=0.2$. The data clearly shows a
peaked structure of the dynamic response functions. Moreover, we show in this
plot the impact of the vertex corrections (compare solid and dashed lines)
which are only quantitative in the considered case as claimed in the
introduction. Fig.~\ref{Fluctuations}d shows the doping dependence of the
characteristic frequency $\omega^{\rm{c,s}}_{0}(\mathbf{q}_\text{max.})$
defined by $\omega^{\rm{c,s}}_{0}(\mathbf{q}_{\text{max.}}) =
\int_{0}^{\infty}\omega\mathrm{Im}\left[\chi^{\rm{c,s}}(\mathbf{q}_{\text{max.}},\omega)\right]d\omega/\int_{0}^{\infty}\mathrm{Im}\left[\chi^{\rm{c,s}}(\mathbf{q}_{\text{max.}},\omega)\right]d\omega$
in both channels. Inside the superconducting region (indicated by the vertical
red dashed lines) the characteristic frequency of the fluctuations are of the
order of $100-200$meV. Moreover, $|\omega^{\rm{c}}_{0}$-$\omega^{\rm{s}}_{0}|$ is
small and minimal for the region of maximum $T_c$. In agreement with our
discussion above we see that an increase of $V$ yields even smaller
$|\omega^{\rm{c}}_{0}$-$\omega^{\rm{s}}_{0}|$ which suggests that charge and spin
contributions to the SC pairing mechanism are cumulative.\\

\begin{figure}[t]
  \includegraphics[width=\columnwidth]{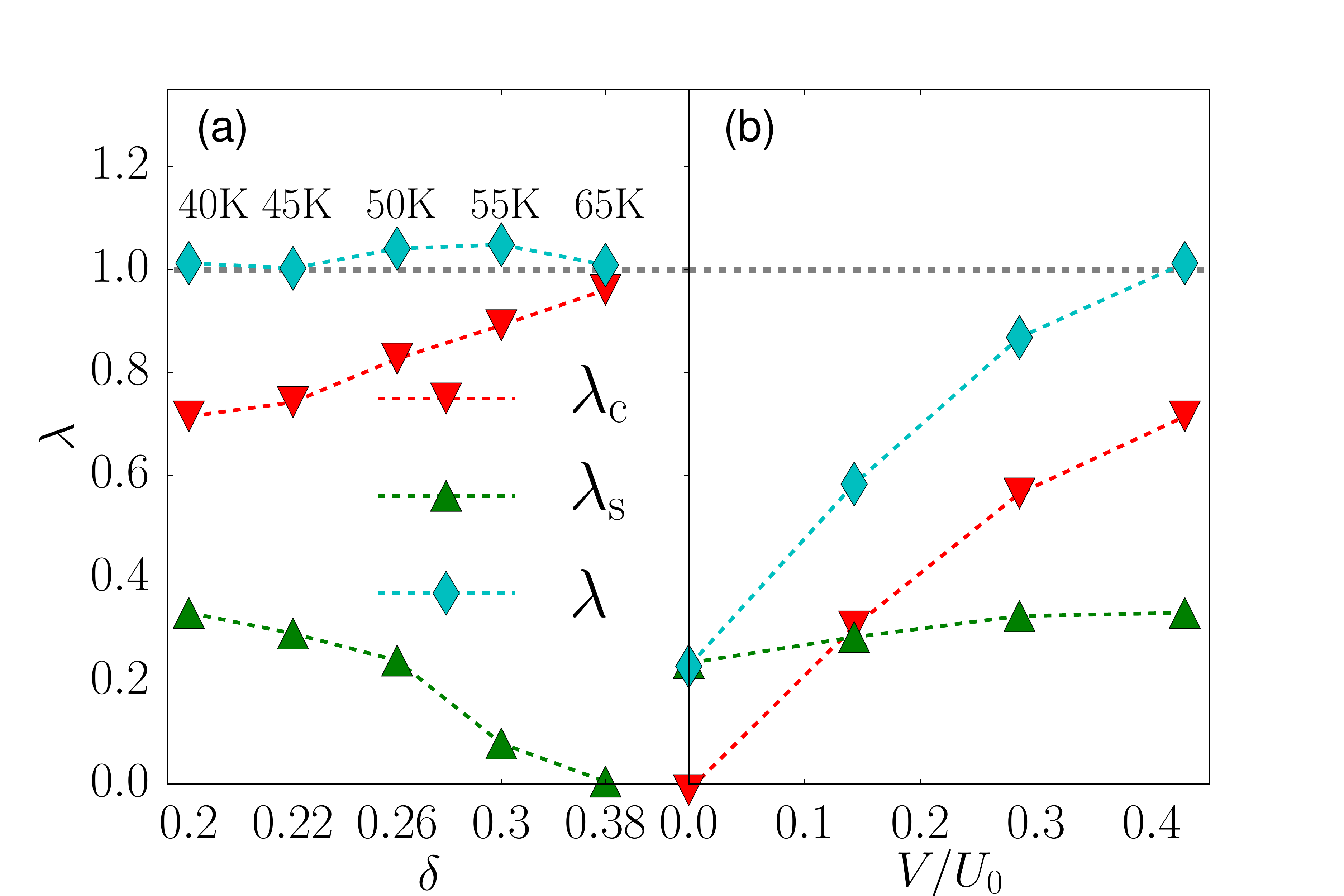}
  \caption{Eigenvalue $\lambda$ of the gap equation Eq.~\eqref{Gap}($\lambda=1$ signals SC transition) for full effective singlet pairing
    interaction $V^{\rm{eff}}_{\mathbf{q},i\nu_{n}}$ (cyan) and charge/spin only channels
    (red/green). (a) Plot for $V=0.3$eV along the SC phase boundary up to
    doping $\delta=0.38$. (b) Plot as a function of $V$ for fixed doping
    $\delta=0.2$ and temperature $T=40$K.}
\label{Cooperation}
\end{figure}

\textit{Separating spin and charge channels in the pairing mechanism --}
In order to disentangle the interplay between charge and spin degrees of
freedom in gap equation~\eqref{Gap}, we solve for $\lambda$ including
contributions from only spin- ($\lambda_s$) and only charge channel ($\lambda_c$), i.e., $V^{\rm{eff}}_{\mathbf{q},i\nu_{n}} =
-3W^{\rm{s}}_{\mathbf{q},i\nu_{n}}$ and $V^{\rm{eff}}_{\mathbf{q},i\nu_{n}} =
W^{\rm{c}}_{\mathbf{q},i\nu_{n}}$ respectively. First, we follow the phase boundary of the
SC phase in the underdoped regime for fixed $V=0.3$eV starting from
$(\delta,T)=(0.2, 40\rm{K})$ up to $(\delta,T)=(0.38, 65\rm{K})$. In
Fig.~\ref{Cooperation}(a) we plot $\lambda$, $\lambda_s$, and $\lambda_c$:
Since we are following the phase transition line, $\lambda\approx
1$. $\lambda_{\rm{c}}$ and $\lambda_{\rm{s}}$ are both smaller than $\lambda$
and $\lambda_{\rm{c}}+\lambda_{\rm{s}} \approx \lambda$ indicating a
cumulative charge and spin contribution for the chiral $d$-wave pairing in the
underdoped regime. The same conclusion can be drawn when the $\lambda$ values
are calculated at the critical doping $\delta_{\rm{c}}=0.2$ as a function of the non-local
interaction $V$ as depicted in Fig.~\ref{Cooperation}(b).  

Our data indicates that overall both spin- and charge fluctuations are
important for the SC phase. As a function of doping, however, we observe that
charge fluctuations become increasingly dominant and $\lambda_{\rm{s}}$
becomes negligible. This effect is reflected in the $V$ dependence of the SC
dome in Fig.\ref{PhaseDiagram} which is stronger at larger dopings. We arrive
at the same conclusions when we analyze the dependence of $\lambda$ on the
choice of the Fierz parameter that defines the charge-to-spin fluctuation
ratio (Suppl. Mat. F).

Let us stress two important points: i) The true long-range character is
crucial in our range of parameters. If only short-range (i.e. nearest-neighbor) interactions are considered charge ordering is overestimated and
long before any SC emerges the system turnes into a charge ordered insulator
as proven by calculations shown in the Suppl. Mat. G. ii)
The degeneracy of $d_{x^{2}-y^{2}}-$ and $d_{xy}-$wave pairing state is
important for the cumulative charge and spin interplay. Since the origin of
this degeneracy is connected to the lattice symmetry group, a different
behavior can be expected for the 2D square lattice (see
Suppl. Mat. H): in the square geometry with relatively large
$V/U_{0}$, the $\mathbf{q}$ dependence of
$\chi^{\rm{c}}(\mathbf{q},i\nu_{n}=0)$ favors $d_{xy}-$pairing symmetry while
$\chi^{\rm{s}}(\mathbf{q},i\nu_{n}=0)$ prefers $d_{x^{2}-y^{2}}-$pairing
symmetry, and the two channels compete with each other.\\

In conclusion we predict the existence of a dome shaped unconventional chiral
$d$-wave superconducting phase for hole-doped triangular lattice systems with
$\propto 1/r$ interactions which could be realized by hole-doping existing
$\alpha$ phase Si(111) ad-atom materials. The analysis of spin and charge
correlation functions reveals that lattice geometry as well as the non-local
interaction are necessary conditions for the emergence of
superconductivity. The nature of the pairing undergoes a crossover from a
combined charge/spin mechanism in the underdoped regime towards a charge
fluctuation dominated one at higher doping. In future studies high hole-doping
levels will be considered in more detail. Here, triplet $f-$wave pairing
symmetry may begin to become important due to the appearance of a disconnected
Fermi surface~\cite{2013_Kuroki_prb}.\\

\subsection{Acknowledgments}
OP and TA are supported by the FP7/ERC, under Grant Agreement
No. 278472-MottMetals. We thank Yi Lu, Alessandro Toschi, Ciro Taranto and Thomas
Schaefer, Daniil Mantadakis for helpful discussions.

\clearpage
\onecolumngrid
\appendix

\begin{center}
\textbf{\large{Supplemental Material}}
\end{center}

\section{Formulation of the non-local interaction $v(\mathbf{q})$ in the lattice model}\label{Sup_vq}
\label{Sup_int}
The long-range interaction in momentum space can be formulated as:
\begin{gather}
 U(\mathbf{q}) = U_{0}+v(\mathbf{q})=U_{0}+V\sum_{i\neq0}\frac{1}{|\mathbf{R}_{i}|/a}e^{i\mathbf{q}\cdot\mathbf{R}_{i}},
\end{gather}
where $a$ is the lattice constant. In order to tackle the convergence problem
given by a Madelung like lattice-sum we follow the ideas of
Ewald and rewrite the sum in terms of a short-range
contribution and a long-range contribution. The long-range contribution can be
obtained analytically, while the short-range contribution is calculated
numerically with a parameter $\eta$ controlling the summation range:
\begin{gather}
v(\mathbf{q})=V\left( \sum_{\substack{\mathbf{R}\in \rm{BL}\diagdown {\{0\}} \\ |\mathbf{R}| <N}} \frac{\rm{erfc}\left(|\mathbf{R}|/\eta\right)}{|\mathbf{R}|} e^{i\mathbf{q}\cdot \mathbf{R}} + \frac{2\pi}{|\mathbf{q}|}\rm{erfc}\left(\frac{|\mathbf{q}|\eta}{2}\right) - \frac{1}{\eta}\frac{2}{\sqrt{\pi}} \right),
\end{gather}
where $\sqrt{N}\ll\eta\lesssim N$ and $N$ is the linear size of the lattice($N=64$ in our calculations). Here we have taken the nearest-neighbor distance $a=1$. The function $\rm{erfc}(x)$ is the complementary of the error function $\rm{erf}(x)$, namely $\rm{erfc}(x)=1-\rm{erf}(x)$. $\rm{BL}$ represents sites in the Bravais lattice.\\

\section{Effects of the vertex $\Lambda^{\rm{imp},\eta}(i\omega_{n},\nu_{n})$}
\label{Sup_lambda}
\begin{figure}[t]
	\includegraphics[width=0.65\textwidth]{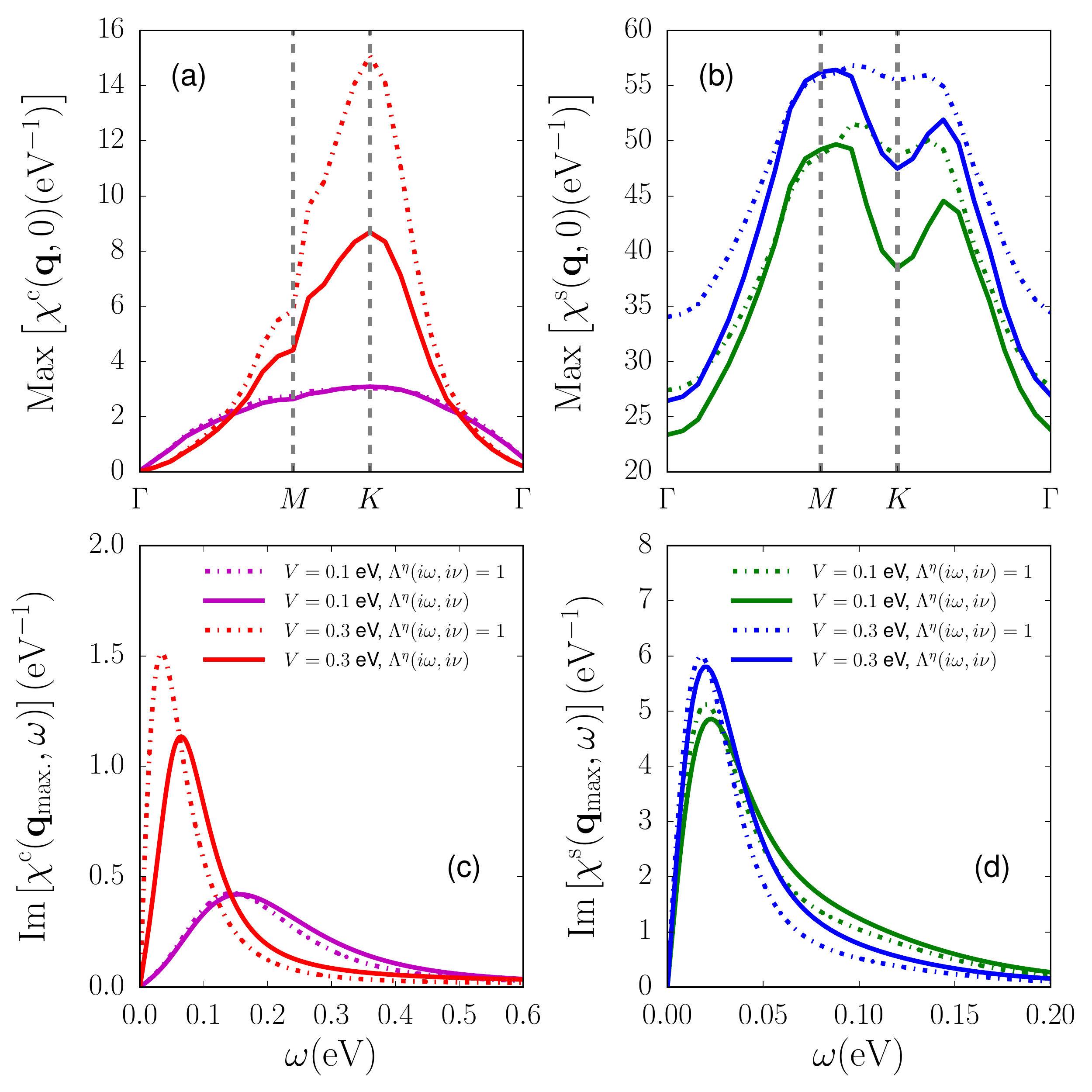}
	\caption{Charge- and spin response functions with (solid lines) and without (dashed lines) vertex corrections for $V=0.1$eV and $V=0.3$eV. Upper panels: Static charge- (a) and spin (b) response function
          along the high symmetry points. Lower panels: Spectrum of charge-
          (c) and spin (d) response function at
          $\mathbf{q}=\mathbf{q}_{\rm{max.}}$ with $\mathbf{q}_{\rm{max.}}$
          being the peak position of the corresponding static response
          function. The shown results were obtained for $U_{0}=0.7$eV,
          $T=116$K and $\delta=0.2$.}
	\label{VertexCorrection}
\end{figure}

Upon increasing the long-range interaction strength from $V=0.1$eV to $V=0.3$eV, the static charge- and spin response functions are enhanced (see
Fig.~\ref{VertexCorrection} (a) and (b)). Simultaneously, their characteristic
frequencies are shifted to lower energies shown in Fig.~\ref{VertexCorrection}
(c) and (d) (note that the data for the dynamic response for $V=0.3$ eV is
shown in the main text figure~\ref{Fluctuations}(c)). Hence, our conclusions
about the $V$ dependence is not compromised by vertex corrections. In
Fig.~\ref{VertexCorrection}(a) we see that
$\Lambda^{\rm{imp},\rm{c}}(i\omega_{n},i\nu_{n})$ partially suppresses the charge
response function. I.e., the critical nearest-neighbor interaction
strength $V_{\rm{c}}$ of metal to charge-ordered phase transition is shifted to
larger values if the three-legged vertex is taken into account. For the spin
response function (Fig.~\ref{VertexCorrection}(b)),
$\Lambda^{\rm{imp},\rm{s}}(i\omega_{n},i\nu_{n})$ slightly suppresses its value and
shifts its maximum closer to $M$. Finally, the $\lambda$ values obtained from
the solution of the gap equation for $V=0.3$eV are actually \emph{increased}
from $0.49$ to $0.52$ as a consequence of the vertex corrections (for
$V=0.1$eV $\lambda$ increases from $0.271$ to $0.274$). This means that
inclusion of vertex corrections leads to even higher values of T$_{\rm{c}}$ which
was found also in another recent TRILEX study for the 2D square lattice
Hubbard model~\cite{2017_thomas_sc_trilex}.

\section{$\rm{Im}\left[\Sigma_{\rm{loc}}(i\omega_{n})\right]$ at different doping levels}
\label{Sup_sigma}
\begin{figure}[h]
	\includegraphics[width=0.55\textwidth]{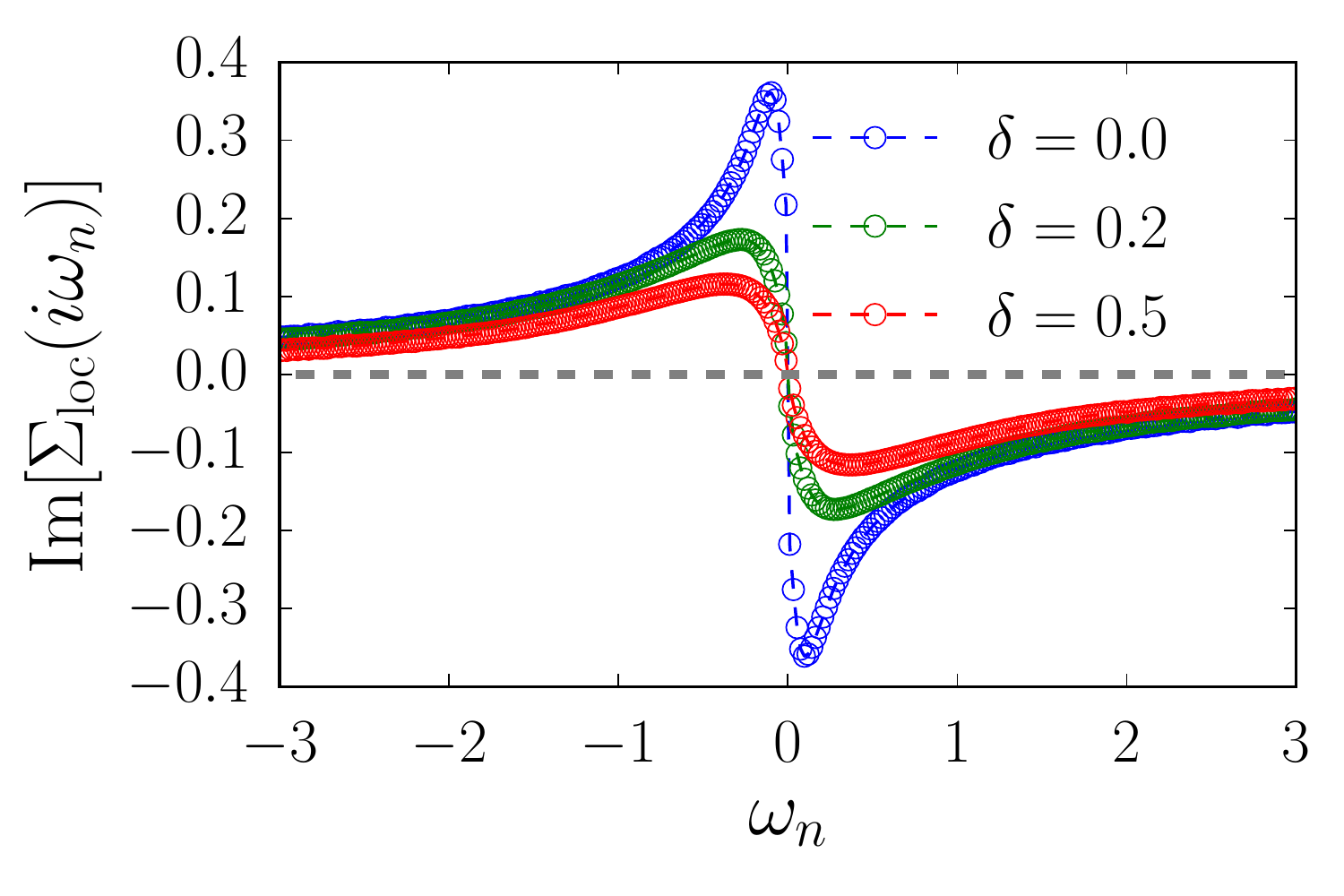}
	\caption{Plot of the imaginary part of the local fermionic self-energy on the Matsubara axis $\rm{Im}\left[\Sigma_{\rm{loc}}(i\omega_{n})\right]$ for several hole doping levels. We show data for fixed $(U_{0},V) = (0.7,0.3)$eV, and $T=40$K.}
	\label{SigmaVsDelta}
\end{figure}

The Fermi liquid character of the normal state above the critical temperature can be seen from the imaginary part of the local fermionic self energy on the Matsubara axis. In Fig.~\ref{SigmaVsDelta} we show $\rm{Im}\left[\Sigma_{\rm{loc}}(i\omega_{n})\right]$ for different doping levels. From the data shown we can estimate the mass enhancement of the correlated quasiparticles $m/m^*=
\left[1-\rm{Im}\left[\Sigma_{\rm{loc}}(i\omega_{0})\right]/\omega_{0}\right]^{-1}=0.047,0.21,0.38$
corresponding to $\delta=0.0,0.2,0.5$ respectively.

\section{Temperature dependence of $\lambda$}
\label{Sup_lvsT}
\begin{figure}[h]
	\includegraphics[width=0.55\textwidth]{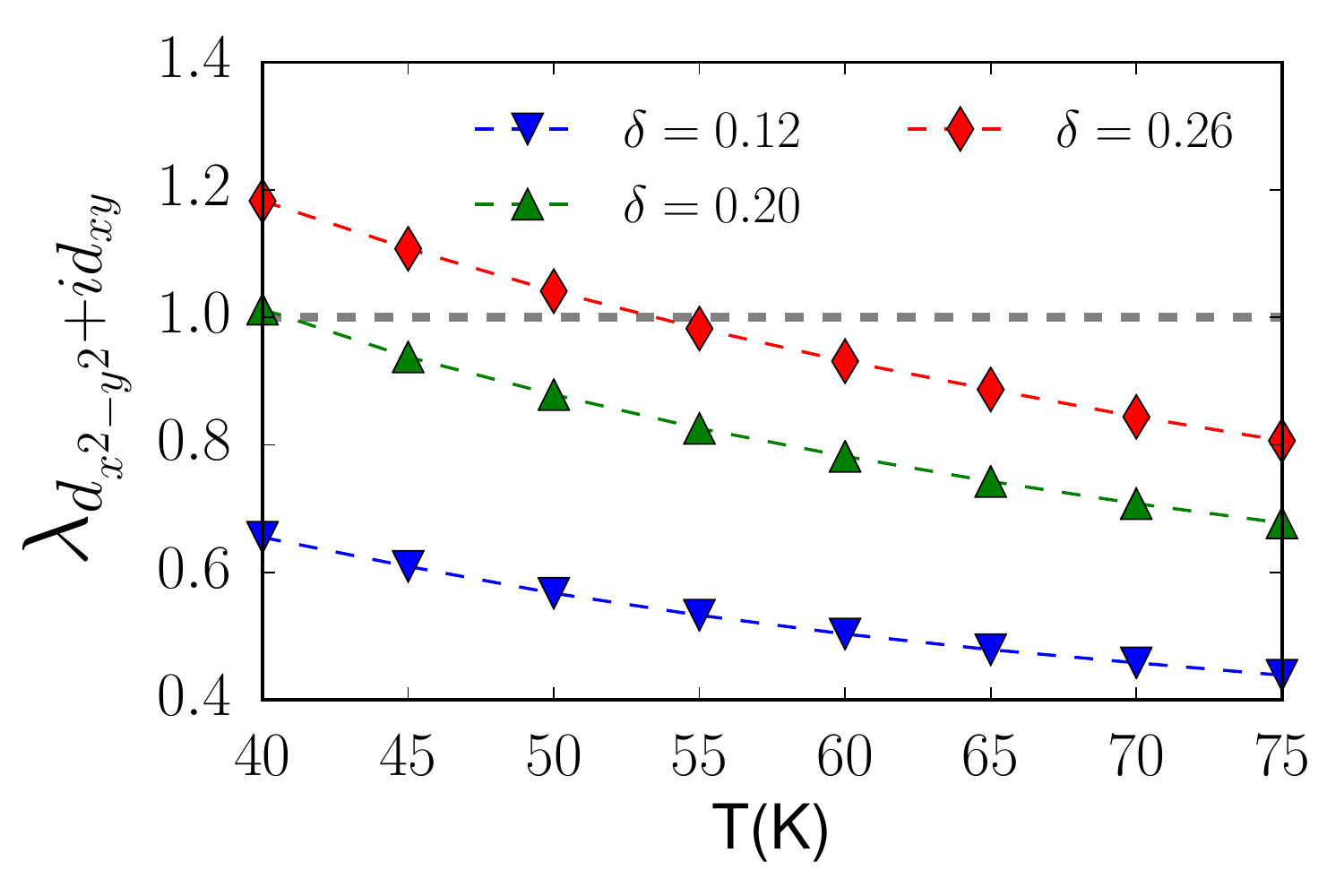}
	\caption{Plot of $\lambda$ as a function of temperature $T$ at different doping levels for fixed $U_0=0.7$eV and $V=0.3$eV.}
	\label{LambdaVsT}
\end{figure}

In Fig.~\ref{LambdaVsT} we plot $\lambda$ as a function of temperature for
different doping levels. The SC instability is indicated by 
$\lambda = 1$ (for instance $\delta=0.2$ at $T\approx 40$K and $\delta=0.26$ at $T\approx 55$K). Please note that no extrapolation of $\lambda(T)$ is needed due to the absence of a magnetically ordered phase, different from the square lattice case~\cite{2017_thomas_sc_trilex}.

\section{Momentum dependence of chiral $d-$wave gap function}
\label{Sup_gap}
\begin{figure}[h]
  \includegraphics[width=0.65\textwidth]{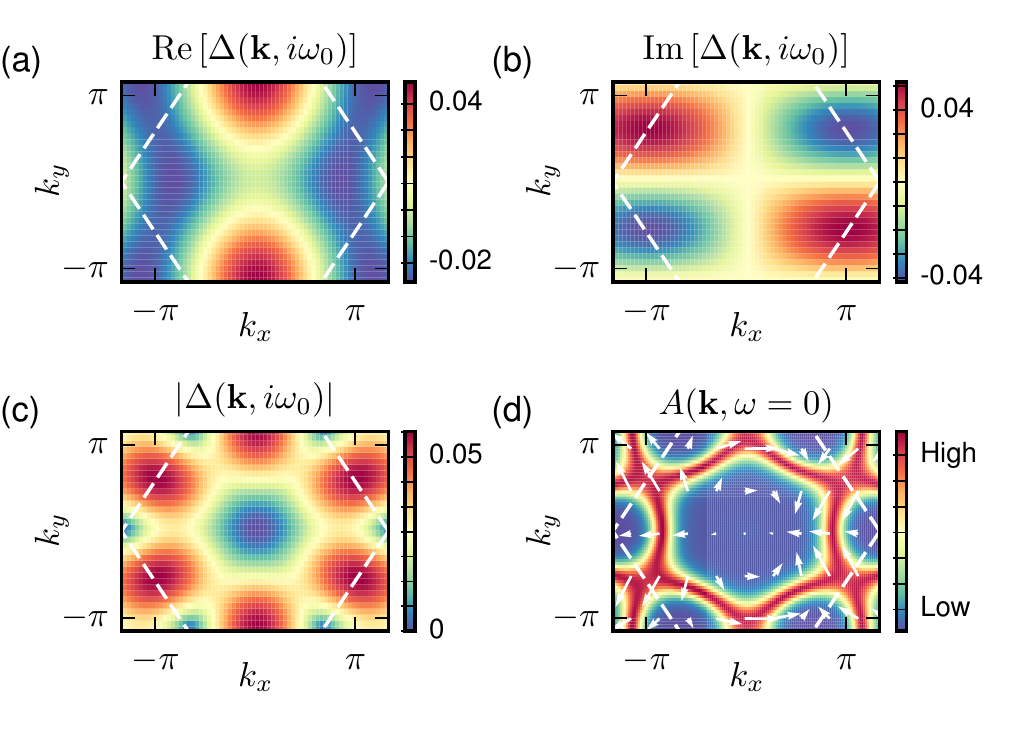}
  \caption{Momentum dependence of the gap function $\Delta_{d+id'}(\mathbf{k},i\omega_{n})$ at
    $\omega_{n}=i\pi T$. (a) $\mathrm{Re}\left[\Delta_{d+id'}\right]$,  (b)
    $\mathrm{Im}\left[\Delta_{d+id'}\right]$ and (c)
    $\left\|\Delta_{d+id'}\right\|$. (d) Plot of the complex gap
    function as vectors $(\rm{Re}\left[\Delta_{d+id'}\right],\rm{Im}\left[\Delta_{d+id'}\right])$ on top of the momentum
    dependent spectral function
    $A(\mathbf{k},\omega=0)=-G(\mathbf{k},\tau=\beta/2)/\pi$. The gap function
    and spectral function were calculated for $T=40$K, $\delta=0.2$ and
    $(U_{0},V)=(0.7,0.3)$eV.}
  \label{GapFunction}
\end{figure}
In Fig.~\ref{GapFunction} we plot the momentum dependence of the chiral
$d-$wave gap function $\Delta_{d+id'}$ obtained from the solution of the gap
equation \eqref{Gap} for the triangular lattice with long-range
interaction. The chiral $d+id'$ superconducting state is a time-reversal
symmetry breaking state with non-trivial topology as can be seen in
Fig.~\ref{GapFunction}(d) from the non-zero winding number($=2$) along the
Fermi surface. This indicates the existence of two edge
states.

\section{Dependence on the charge to spin ratio}
\label{Sup_alpha}
\begin{figure}[h]
	\includegraphics[width=0.55\textwidth]{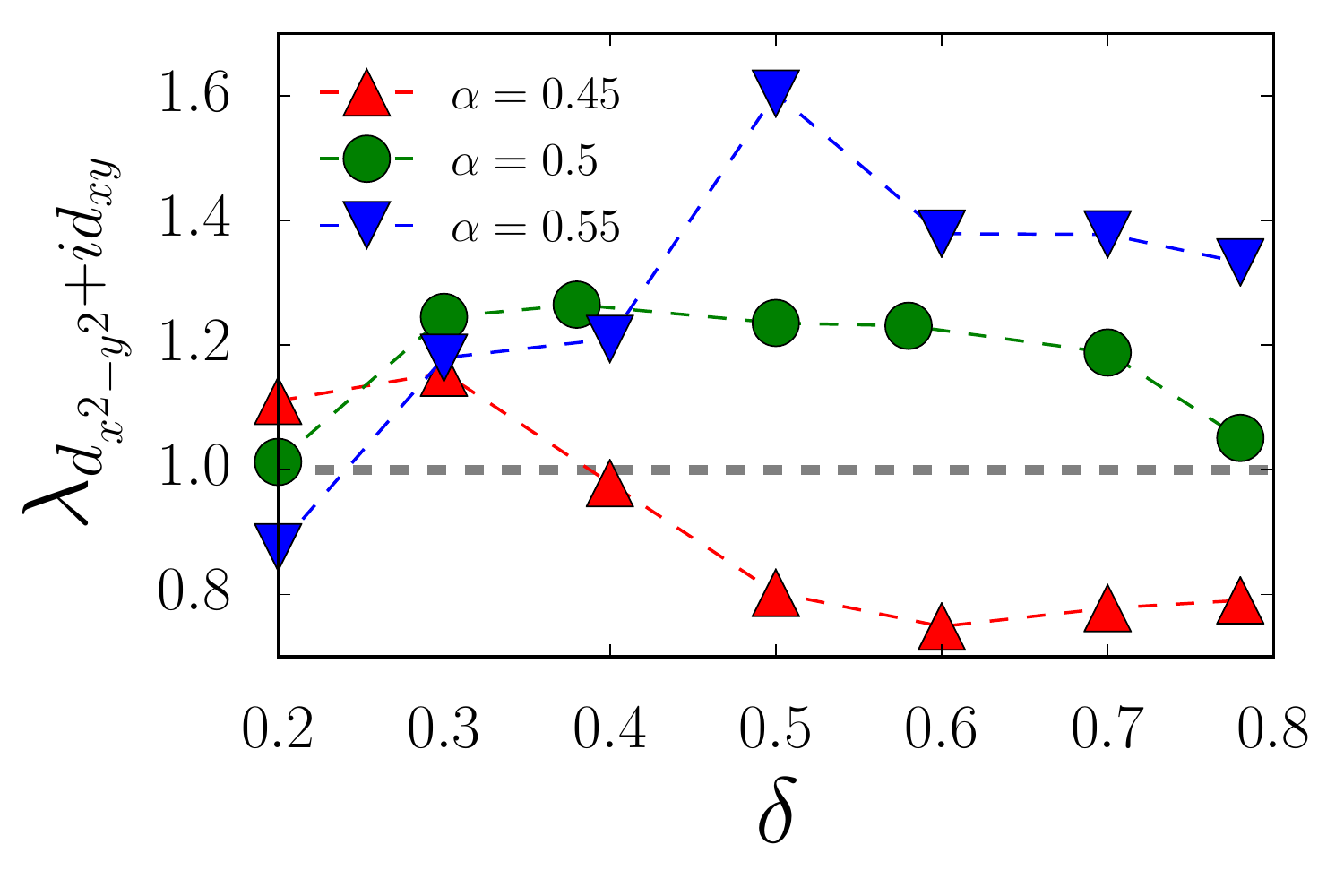}
	\caption{Chiral $d-$wave $\lambda$ values corresponding to different values of the Fierz parameter $\alpha$ as a function of doping. The shown data was obtained at $T=40$K, $U_{0}=0.7$eV, and $V=0.3$eV.}
	\label{AlphaDependec}
\end{figure}
As mentioned in the main text, the ratio of the bare interaction in charge-
and spin channels may be parametrized by $\alpha$, i.e.,
\begin{gather}
	U^{\rm{c}}(\mathbf{q})=(3\alpha-1)U_{0}+v(\mathbf{q}),
	U^{\rm{s}}=(\alpha-2/3)U_{0}
\end{gather}
for Heisenberg decoupling~\cite{2015_PRB_trilex_thomas}. The TRILEX results
depend a priori on the choice of the Fierz parameter $\alpha$. In the
following, we show that our conclusions are robust with respect to the choice
of $\alpha$. While there is a sizable dependence of the $\lambda$ values on
$\alpha$, this dependency leads only to a quantitative shift of the boundary
of the SC phase but SC is never suppressed.

Since $\alpha$ controls the contributions from charge- and spin fluctuations
to the SC pairing glue, we can exploit the dependence of the results on
$\alpha$ as an indicator of their respective role in the emergence of SC. As
shown in Fig.~\ref{AlphaDependec}, for comparatively small doping
($\delta\sim0.2$) $\lambda$ is increased by decreasing $\alpha$
(i.e. emphasizing spin fluctuations). This indicates that at small doping spin
fluctuations are the main contributor to the emergence of
superconductivity. At large doping ($\delta\geq 0.5$), in contrast, $\lambda$
is increased by increasing $\alpha$ (emphasizing the charge channel), which
indicates once more that charge fluctuations are key for the emergence of
superconductivity at large doping. Finally, for intermediate doping
($\delta\in(0.3,0.42)$), the largest $\lambda$ value is found for
$\alpha=0.5$, indicating that in this region charge- and spin fluctuations
contribute ``cumulatively'' to the SC instability. While it is hard to
further disentangle the cross influence of charge and spin fluctuations in the
self-consistent solution, the insights from the $\alpha$ dependence support the
picture of a cooperative (or additive) spin-charge pairing mechanism as
discussed in the main text.

\section{Long-range versus short-range non-local interaction}\label{Sup_short}
\label{Sup_shortrange}
\begin{figure}[h]
  \includegraphics[width=0.65\textwidth]{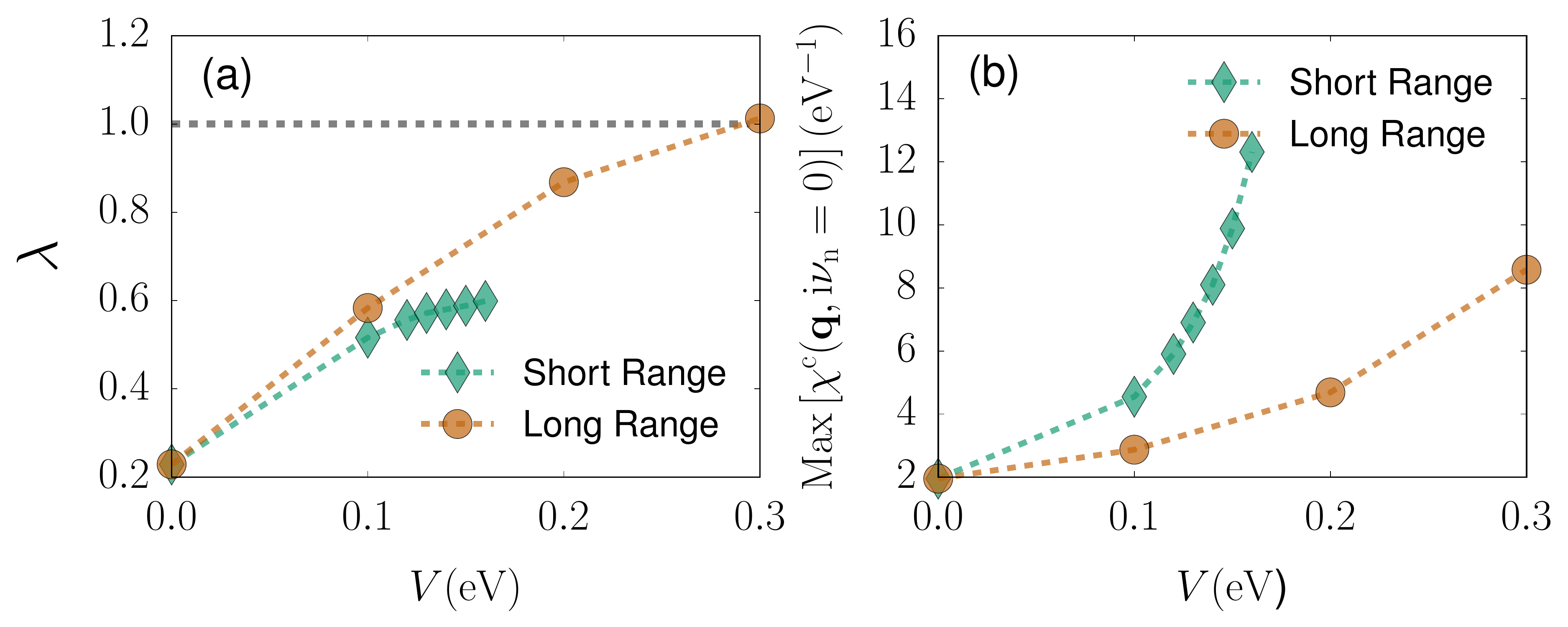}
  \caption{Comparison of short- and long-range interaction. Here $V$
    represents the nearest-neighbor interaction for short-range interaction
    and the $1/r$ prefactor for long-range interaction. (a) $\lambda$ values
    as a function of $V$ for short- (diamond) and long-range interaction
    (circle). (b) Maximum of static charge susceptibility as a function of
    $V$. The parameters are $U_{0}=0.7$eV, $T=40$K and $\delta=0.2$.} 
  \label{SubShort}
\end{figure}
We now show that it is not possible to obtain the same phase diagram (in
particular the superconducting phase) with non-local but short-range
(e.g. nearest-neighbor) interaction. In Fig.~\ref{SubShort}(a) we show
$\lambda$ as a function of $V$ for the short-range (diamonds) and long-range
(circles) interaction. Please note that $V$ denotes the strength of the $1/r$
tail when long-range interactions are considered while it represents
nearest-neighbor interactions only for the short-range version. We not only
observe a downturn of $\lambda$ upon increasing $V$ but, most importantly, a
dramatic increase in the associated charge response functions indicating a
second order phase transition to a charge ordered phase
(Fig.~\ref{SubShort}(b)). Hence, when only nearest-neighbor interaction is
considered, a charge order instability will occur long before superconducting
fluctuations become sizable. In the case of true long-range interactions, the
situation is quite different and charge (and spin) fluctuations are enhanced
but remain finite up to the point of $\lambda=1$.

\section{Comparison to the square lattice}
\label{Sup_squarelat}
\begin{figure}[h]
  \includegraphics[width=0.75\textwidth]{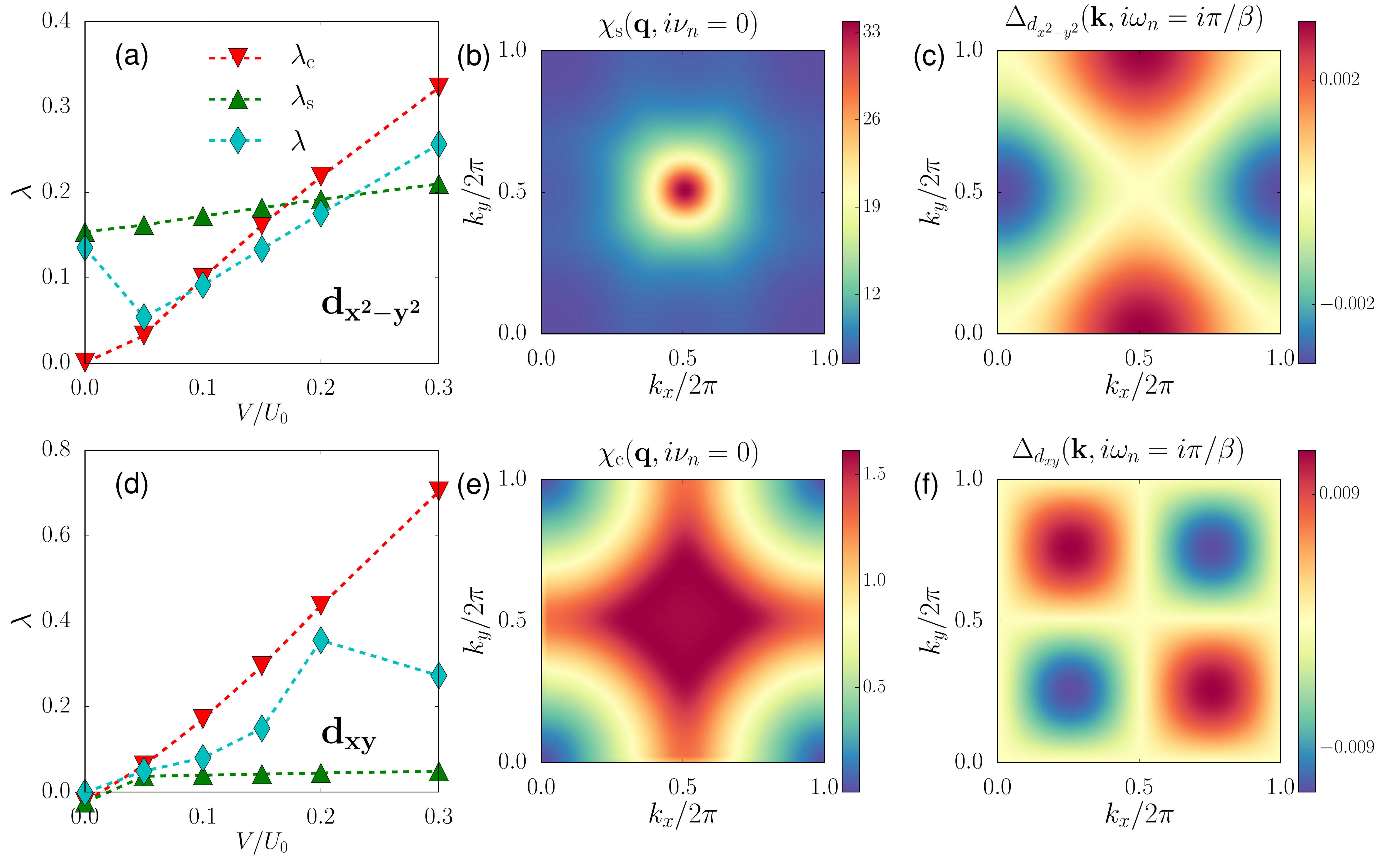}
  \caption{Simplified TRILEX results for the square lattice with long-range
    interaction. The parameters are chosen as $t=-0.25$eV and $t'=-0.2t$
    corresponding to nearest-neighbor and next-nearest-neighbor hopping
    integrals, on-site interaction $U_{0}=2.0$eV and fixed temperature 
    $T=290$K. Upper panel: (a) $\lambda$ values for $d_{x^{2}-y^{2}}$
    pairing symmetry as a function of long-range interaction strength $V$
    computed with charge(down triangular), spin(up triangular) and
    combined(diamond) contributions; (b) static spin response function;
    (c) solved $d_{x^{2}-y^{2}}$ gap function at
    $i\omega_{n}=i\pi/\beta$. Here hole doping level is $\delta=0.2$, and
    $V=0.0$eV for  (b) and  (c).  Lower panel:(d) $\lambda$ values
    for $d_{xy}$ pairing symmetry as function of long-range interaction
    strength $V$ computed with charge(down triangular), spin(up triangular)
    and combined(diamond) contributions; (e) static charge response
    function; (f) solved $d_{xy}$ gap function. Hole doping level is
    fixed at $\delta=0.5$ and $V=0.6$eV for (e) and  (f).} 
\label{SubSquare}
\end{figure}
As discussed in the main text, the charge/spin pairing mechanism of our chiral
SC instability depends crucially on the degeneracy of the $d_{x^{2}-y^{2}}-$
and $d_{xy}$ pairing state. This is the case for the triangular lattice where
both states belong to the same irreducible representation (E$_2$). For
different lattice geometries where $d_{x^{2}-y^{2}}-$ and $d_{xy}$ pairing
states are not degenerate the interplay between charge- and spin fluctuation
for the SC instability can be qualitatively different from our model. As an
important example we mention the 2D square lattice for which
$d_{x^{2}-y^{2}}-$ and $d_{xy}$ belong to different irreducible
representations B$_1$ and B$_2$, respectively. In Fig.~\ref{SubSquare}(a)  and
(d) we show the corresponding $\lambda$ values obtained in the square lattice
as a function of the long-range interaction $V$ for both $d_{x^{2}-y^{2}}-$
and $d_{xy}$ pairing symmetries. With the same separation of channel
contribution as performed in the main text, we see a qualitative difference in
the behavior of $\lambda$: on the triangular lattice, $\lambda$ is larger than
$\lambda_{\rm{c}}$ and $\lambda_{\rm{s}}$, while on square lattice $\lambda$
is in between or smaller than $\lambda_{\rm{c}}$ and $\lambda_{\rm{s}}$.

In order to disentangle the singlet-pairing interaction in the
particle-particle channel into charge- and spin contributions we use
\begin{gather}\label{decoupling}
	V^{\rm{eff}}(\mathbf{q},i\nu_{n}) = W^{\rm{c}}(\mathbf{q},i\nu_{n})  - 3W^{\rm{s}} (\mathbf{q},i\nu_{n}) 
	= \frac{U^{\rm{c}}(\mathbf{q})}{1-U^{\rm{c}}(\mathbf{q})P^{\rm{c}}(\mathbf{q},i\nu_{n}) } - 3\frac{U^{\rm{s}}}{1-U^{\rm{s}}P^{\rm{s}}(\mathbf{q},i\nu_{n}) } \\ \nonumber
	= U(\mathbf{q}) +\underbrace{\frac{U^{\rm{c}}(\mathbf{q})P^{\rm{c}}(\mathbf{q},i\nu_{n}) U^{\rm{c}}(\mathbf{q})}{1-U^{\rm{c}}(\mathbf{q})P^{\rm{c}}(\mathbf{q},i\nu_{n}) }}_{\textbf{charge},-} 
	\underbrace{-3\frac{U^{\rm{s}}P^{\rm{s}}(\mathbf{q},i\nu_{n}) U^{\rm{s}}}{1-U^{\rm{s}}P^{\rm{s}}(\mathbf{q},i\nu_{n}) }}_{\textbf{spin},+},
\end{gather}
with $U^{\rm{c}}(\mathbf{q}) = \frac{U_{0}}{2}+v(\mathbf{q})$. The $+$($-$)
denotes the positive/negative contribution from each channel ($P^\text{c/s}<0$
in our parameter range). We denote the typical pairing-scattering momentum for
charge- and spin channel as $Q_{\rm{c}}$(Fig.~\ref{SubSquare}(e)) and $Q_{\rm{s}}$(Fig.~\ref{SubSquare}(b)) respectively
(i.e. momenta where $\chi_{\text{c/s}}$ are maximal). In order to find a large
$\lambda$ value when solving Eq.~\eqref{Gap} $\Delta(\mathbf{k},i\omega_{n})$ should not change sign
for scattering with $Q_{\rm{c}}$ in the charge channel, while it should change
sign when scattering with $Q_{\rm{s}}$. Hence, when spin fluctuations dominate, the $d_{x^{2}-y^{2}}$(Fig.~\ref{SubSquare}(c)) pairing symmetry
is favorable in the $d-$wave singlet pairing and charge fluctuations
contribute destructively. Vice versa, when charge fluctuations dominate, $d_{xy}$(Fig.~\ref{SubSquare}(f)) symmetry will be the favored.

\noindent\makebox[\linewidth]{\resizebox{0.3333\linewidth}{2pt}{$\bullet$}}\bigskip

\end{document}